\documentclass[12pt]{article}
\usepackage{amsbsy,graphicx,latexsym}

\newcommand{\beq}{\begin{equation}}
\newcommand{\eeq}{\end{equation}}
\newcommand{\bea}{\begin{eqnarray*}}
\newcommand{\eea}{\end{eqnarray*}}
\newcommand{\beaq}{\begin{eqnarray}}
\newcommand{\eeaq}{\end{eqnarray}}

\begin{document}
\begin{flushright}hep-th/0205193\end{flushright} 

\vspace{10mm}
\centerline{\Large \bf Unitarity in space-time noncommutative field theories}
\vskip 1cm
\centerline{\large Chaiho Rim$^{1}$ and Jae Hyung Yee$^{2}$}
\vskip 1cm 
\centerline{\it $^{1}$ Department of Physics, Chonbuk National University}
\centerline{\it Chonju 561-756, Korea}
\centerline{\it rim@mail.chonbuk.ac.kr}
\vskip .5cm
\centerline{\it $^{2}$ Institute of Physics and Applied Physics, Yonsei University}
\centerline{\it Seoul 120-749, Korea}
\centerline{\it jhyee@phya.yonsei.ac.kr}
\vskip 2cm

\centerline{\bf Abstract}
\vskip 0.5cm
\noindent 
In non-commutative field theories conventional wisdom is that 
the unitarity is non-compatible 
with the perturbation analysis when time is involved 
in the non-commutative coordinates.
However, as suggested by Bahns et.\ al.\ recently,
the root of the problem lies in the improper definition 
of the time-ordered product. 
In this article, functional formalism of S-matrix is explicitly constructed 
for the non-commutative $\phi^p$ scalar field theory 
using the field equation in the Heisenberg picture
and proper definition of time-ordering. 
This S-matrix is manifestly unitary.  
Using the free spectral (Wightmann) function 
as the free field propagator, 
we demonstrate the perturbation obeys the unitarity, 
and present the exact two particle scattering amplitude 
for 1+1 dimensional non-commutative 
nonlinear Schr\"odinger model. 
\newpage

\section{Introduction}

Quantum field theory on noncommutative spacetimes arises typically 
in restrictive phase space~\cite{nc} and has some applications 
in condensed matter physics such as in quantum Hall effect~\cite{qhe}.
This formalism has much more interesting features 
if the non-commuting coordinates involve time, 
i.e. non-commuting space-time. 
The framework of this noncommutative spaces can implement
the possible deviations from the smoothness of spacetime at small distances 
and results in a modification of uncertainty relations for
spacetime coordinates~\cite{sw}. 

Despite this facinating possibility in space-time non-commutative field theories, 
in the perturbative field theories~\cite{ncpert}
it is asserted that the theories possess a serious problem, i.e., 
the lack of unitarity~\cite{stnc} and there are some atempts to cure this problem 
such as in the Hamiltonian picture~\cite{gkl}.  

Contrary to this view, Bahns et.\ al.~\cite{bahns} recently 
pointed out that this unitarity problem is not inherent 
in the non-commutative field theories but rather 
due to the ill-defined time-ordered product expansion. 

In this article we elaborate on this view. In section 2, 
we present the S-matrix explicitly in the functional form and show
how unitarity problems are cured. In terms of perturbative loop correction,
the same result is presented in section 3. 
As a further concrete example, we present exact 2-particle 
scattering amplitude for the non-commutative version of the integrable 
non-linear Schr\"odinger model in 1+1 dimension.  

\section{S-matrix}
\noindent             

Quantum field theory on the noncommutative spacetime can be  
constructed into a nonlocal field theory on a commutative spacetime,
using  $\star$-product of fields. 
One of the convenient $\star$-product representations is the Moyal product, 
\beq 
f\star g\, (x)=
e^{\frac{i}{2}{\partial}_x \wedge 
{\partial}_y}f(x)g(y)|_{y=x} \,
\eeq
where $a \wedge b = a_\mu \theta^{\mu\nu} b_\nu$. 
$\theta^{\mu\nu} $ is an antisymmetric c-number 
representing the space-time non-commutativeness, 
$i \theta^{\mu\nu} =[x^\mu , x^\nu]$.
This Moyal product makes the kinetic term of the action 
the usual field theory, 
and allows the conventional perturbation 
with the proper vertex correction 
corresponding the nonlocal interaction~\cite{ncpert}. 

We adopt a real scalar field theory for simplicity.  
The Lagrangian constitutes of the free part 
and interacting part. The interaction Lagrangian in $D-1$ space 
is given  as
\beq
L_I (t) = -\frac{g}{p!} \,\int d^{D-1}x \,\, \frac12\,  
( \phi_\star^p (x,t)+ {\rm h.c. })
\eeq
where $g$ is a coupling constant.
$\phi_\star^p = \phi \star \phi \star \cdots \star \phi$ 
is the non-commutative version of $\phi^p$ theory where 
p is a positive integer. 
We make the action manifestly hermitean by adding 
the hermitean conjugate part.
 
To construct the S-matrix, 
one assumes the out-going field 
satisfy the in-coming free field commutator relation
\beq
[\phi_{\rm in} (x), \phi_{\rm in} (0)] = i \bigtriangleup \!(x)\,
\label{commutator}
\eeq
so that the in- and out- fields are related by 
\beq
\phi_{\rm out} = S^{-1}\, \phi_{\rm in} \,\,S \,.
\label{S-relation}
\eeq
This relation is not, however, automatically satisfied.
It is demonstrated in \cite{non-local} 
that non-local field theories may not respect the assumption. 
The out-field commutator relation need be checked
to be consistent. 

We quantize the field using the Heisenberg picture~\cite{yang-feldman}. 
The field at arbitrary time can be obtained from the field equation 
\beq
(\Box + m^2 ) \,\phi (x) = \xi( \phi(x))
\label{h-eq}
\eeq
where $\xi$ is the functional of fields, derived from
the interaction Lagrangian
\beq
\xi(\phi(x)) \equiv \frac{\delta}{\delta \phi(x)} \int dt L_I (t)
 = - \frac{g}{(p-1)!}\,\, \phi_\star^{p-1}(x) \,.
\eeq
Its solution is given using the retarded progator
$\bigtriangleup_{\rm ret}  (x) = -\theta(x^0 )\bigtriangleup\! (x) $ 
(advanced propagator  
$ \bigtriangleup_{\rm ad} (x) =\theta(-x^0 )\bigtriangleup\! (x)$ ), 
\beaq
\phi (x) 
&=& \phi_{\rm in}(x) + \bigtriangleup_{\rm ret} \circ \xi (\phi(x))
\nonumber\\
&=&\phi_{\rm out}(x) + \bigtriangleup_{\rm ad} \circ \xi(\phi(x))\,,
\eeaq
where $\circ$ denotes the convolution, 
$\bigtriangleup_{\rm ret} \circ \xi (x) =
\int d^D y \,\, \bigtriangleup_{\rm ret} (x-y) \,\,\xi(y) $.

Now the out-field can be put iteratively in terms of the in-field,
\beq
\phi_{\rm out}(x) 
= \phi_{\rm in}(x) - \bigtriangleup \circ \xi(\phi(x))\,, 
\label{phiout}
\eeq
if $\phi$ is written as 
$ \phi  = \phi_0 + \phi_1 + \phi_2 \cdots $
where $\phi_n$ represents the order of $g^n$ contribution.
A few explicit solutions of $\phi_n$'s are given as 
\beaq 
\phi_0 (x) &=& \phi_{\rm in} (x)
\nonumber\\
\phi_1 (x) &=&  - \frac{g}{(p-1)!} \bigtriangleup_{\rm ret} 
\circ \,\,\phi_{0\,\star}^{(p-1)} (x)
\nonumber\\
\phi_2 (x) &=& - \frac{g}{(p-1)!} \bigtriangleup_{\rm ret} 
 \circ \,\,
(\phi_1\star\phi_{0\,\star}^{(p-2)}
+ \phi_0\star\phi_1\star\phi_{0\,\star}^{(p-3)} + \cdots
+ \phi_{0\,\star}^{(p-2)}\star\phi_1 ) (x)\,.
\nonumber
\eeaq
As $x^0 \to \infty$ the fields $\phi(x)$ reduces 
to the out-field $\phi_{\rm out}$ 
and $\Delta_{\rm ret} (x) \to - \Delta$ in consistent 
with Eq.~(\ref{phiout}). 

One can check after some tedious calculation that
the out-field $\phi_{\rm out}(x)$ in Eq.~(\ref{phiout}) 
does satisfy the in-field commutator Eq.~(\ref{commutator})
order by order.
Remarkably, the $\star$-product of the action 
does not affect the commutation relation.
This justifies the assumption of the unitary S-matrix 
between in-fields and out-fields
in contrast with general nonlocal theories 
found in \cite{non-local}.

With the notation $S = e^{i\delta}$,
the out-field would be written as
\beq
\phi_{\rm out} = S^{-1} \,\phi_{\rm in} \,\,S 
=\phi_{\rm in} + [\phi_{\rm in}, i \delta ]
+ \frac12 [[\phi_{\rm in} , i \delta ], i\delta ]
+ \cdots \,.
\label{S-delta}
\eeq
The first order term in $g$ results in the equation,
$ [\, \phi_{\rm in}, i \delta \,]
=  - \bigtriangleup \circ \,\,
\xi(\phi_{\rm in}(x)) $,
and determines $\delta$ to the first order in $g$  as
\beq
\delta =  \int_{-\infty}^{\infty} dt\, 
 L_I (\phi_{\rm in}(t))
+ O(g^2)\,.
\eeq
Higher order solutions requires the time-ordering 
as in the ordinary field theory.
However, the $\star$-product introduces a subtlety 
in the time-ordering and a consistent unitary S-matrix
is given as
\beaq    
S &=&  1 +i \int_{-\infty}^\infty dt \,{\cal F}_1 \,
\Big( V (\phi_{\rm in}(t))\Big)
+ i^2 \int_{-\infty}^\infty \!\int_{-\infty}^\infty 
dt_1 dt_2 \,
{\cal F}_{12}\, \Big(\theta_{12} V (\phi_{\rm in}(t_1)) 
V (\phi_{\rm in}(t_2))\Big)
\cdots \nonumber \\ 
&& \quad + 
i^n \int_{-\infty}^\infty \!
\cdots \int_{-\infty}^\infty d t_1 \, \cdots \,d t_n 
{\cal F}_{1 2\cdots n}\,
\Big( \theta_{12 \cdots n} 
V (\phi_{\rm in} (t_1) )\cdots V (\phi_{\rm in} (t_n) )
\Big) + \cdots   \,.                                
\label{s-matrix}
\eeaq
$V (\phi_{\rm in}(t)) $  
is interaction Lagrangian before $\star$-product,  
\bea    
V (\phi_{\rm in}(t)) 
\equiv -\frac{g}{p!} \,\int d^{D-1}x \,\, 
\phi_{\rm in}^{p} (x,t) \,,
\eea
and the time-ordering is given in terms of the step function,
\bea
\theta_{12\cdots n} = \theta (t_1 -t_2) \,
\theta (t_2- t_3) \cdots \theta(t_{n-1} -t_n)\,.
\eea
$\star$-operation ${\cal F}_{1 2\cdots n}$ introduces the 
$\star$-product to the actions
\beq  
{\cal F}_{1 2\cdots n}
\Big( V (t_1) V(t_2) \cdots V (t_n) \Big) = 
L_I (t_1) L_I(t_2) \cdots L_I (t_n) \,,
\eeq
whose operation is independent of 
the permutation of the action.
In the presence of the step-function, we assume 
a minimal realization. For example, explicitly we put 
\bea
{\cal F}_{x y}
\Big( \theta(x^0-y^0) \phi^p (x) \phi^p (y)  \Big) 
= {\cal F}_x \,{\cal F}_y
\Big( \theta(x^0 -y^0) \,
\phi(x_1) \cdots \phi(x_p)\,
\phi(y_1) \cdots \phi(y_p)
\Big) \Big|_{x_i = x\,, y_i=y}
\eea
where 
${\cal F}_x \equiv
 \cos \Bigg(\frac 12 \Big(\partial_{x_1} 
\wedge (\partial_{x_2}+ \cdots + \partial_{x_p})
+ \partial_{x_2} \wedge 
(\partial_{x_3} + \cdots \partial_{x_p})
+\cdots   + \partial_{x_{p-1}} \wedge 
\partial_{x_p} \Big)\Bigg)$
and $\theta(x^0 -y^0)$ is put to $\theta(x_i^0 -y_j^0) $
in the presence of the spectral function 
$\Delta(x_i^0 -y_j^0) $. This operation is done 
explicitly below Eq.~(\ref{outfieldcheck}) 
and Eq.~(\ref{one-loop}) in the next section.

Introducing the time-ordering with $\star$-product,
\beq
T_{\star} \{V(t_1) V(t_2)\} = 
{\cal F}_{12} \Big( \theta_{12} \, V(t_1) V (t_2)
+ \theta_{21} \, V(t_2) V (t_1)\Big)\,.
\eeq 
we can put the S-matrix as 
\beaq
S &=& \sum_{n=0}^\infty \frac{i^n}{n!}
\int_{-\infty}^\infty dt_1 
\cdots \int_{-\infty}^\infty dt_n
T_{\star} \{V (\phi_{\rm in}(t_1))
\cdots  V (\phi_{\rm in}(t_n) \}
\nonumber\\
&\equiv& T_{\star} 
\exp \Big( i\int_{-\infty}^\infty dt \,
V(\phi_{\rm in}(t) \Big)\,.
\eeaq

One can check order by order that 
this S-matrix is unitary $ S^{-1} = S^\dagger$
and reproduces the in- and out-field relation 
Eq.~(\ref{phiout}).
We present here the sketch of the proof of unitarity of the S-matrix 
up to the order of $g^2$.
The higher order proof goes similarly with 
the ordinary perturbation case since in this proof
only the time-ordering matters irrespective of the 
$\star$-operation. 
The unitarity of the S-matrix in Eq.~(\ref{s-matrix}) is 
proved if the following identity is satisfied:
$A_2 + A_2^\dagger = A_1^\dagger A_1 = A_1 ^2 $
where 
\bea
A_1 = \int_{-\infty}^\infty dt_1 {\cal F}_{1} ( V_1) \,,
\qquad 
A_2 = \int_{-\infty}^\infty \!\int_{-\infty}^\infty 
dt_1 dt_2 {\cal F}_{12} \Big( \theta_{12} V_1 V_2 \Big)\,.
\eea
The proof goes as follows:
\beaq
A_2 + A_2^\dagger &=&
\int_{-\infty}^\infty \!\int_{-\infty}^\infty 
dt_1 dt_2 {\cal F}_{12} \Big( \theta_{12} 
(V_1 V_2+ V_2 V_1) \Big)
\nonumber\\
&=&
\int_{-\infty}^\infty \!\int_{-\infty}^\infty 
dt_1 dt_2 {\cal F}_{12} \Big( (\theta_{12} +\theta_{21}) 
V_1 V_2 \Big)
\nonumber\\
&=&
\int_{-\infty}^\infty dt_1 {\cal F}_{1} ( V_1 ) 
\int_{-\infty}^\infty dt_2 {\cal F}_{1} ( V_2 )\,
= A_1^\dagger A_1 
\eeaq
where we use the change of variables to get the second line
and the identity $\theta_{12}+ \theta_{21} =1$ 
for the last line.

On the other hand, the out field 
is obtained from the S-matrix relation:
\beaq
S^\dagger \phi_{\rm in}(x)  S 
&=& \phi_0 (x) 
+ i\int dy\, \Big(\phi_0(x) A_1(y) - A_1(y) \phi_0(x)\Big) 
\nonumber\\
&&\quad
+ i^2 \int dy_1 dy_2 \,\Big(\phi_0(x) A_2(y_1,y_2) 
- A_1(y_1)^\dagger \phi_0 A_1(y_2) 
+ A_2(y)^\dagger \phi_0(x) \Big) 
+ O(g^3) \nonumber \\
&=& \phi_0 + i\int dy_1 \,{\cal F}_1\,
\Big(  [\phi_0(x), V(y_1)] \Big)
\nonumber\\ 
&&\qquad
+i^2 \int dy_1 dy_2 \,{\cal F}_{12}\,
\Big(  \theta_{12} [[\phi_0(x), V(y_1)], V(y_2)] \Big)
+ O(g^3) \,.
\label{outfieldcheck}
\eeaq
It is clear that the out field relation in Eq.~(\ref{phiout})
up to the order $g^2$ is reproduced in 
Eq.~(\ref{outfieldcheck}) 
if one uses the commutation of the fields 
$[[\phi_0(x), V(y_1)], V(y_2)]$
and the time-ordering step function $\theta_{12}$ 
before peforming the $\star$-operation.

We give some comments on other approaches of finding the 
unitary S-matrix. 
First, one may start with the time-ordering outside the 
$\star$-operation as in \cite{bahns},
then one needs higher derivative corrections, which will 
finally reproduce the above S-matrix Eq.~(\ref{s-matrix}).
For example, we put $A_2 =a_2 + i c_2$ at the order $g^2$, 
\beaq
a_2 =\int_{-\infty}^\infty \!\int_{-\infty}^\infty 
\theta_{12} {\cal F}_{12} \Big( V_1 V_2 \Big)\,,\qquad
i c_2 =-\frac12 
\int_{-\infty}^\infty \!\int_{-\infty}^\infty 
\Big(\theta_{12}\, {\cal F}_{12}
-{\cal F}_{12} \,\theta_{12}\Big) 
\Big( \,[V_1, V_2]\Big) \,.
\eeaq
$a_2$ is the ordinary time-ordered one and 
$a_2 + a_2^\dagger =A_1^2$. The correction term $c_2$ 
satisfies the relation $c_2 = c_2^\dagger$
(note that the $\dagger$ operation is applied 
to the field $\phi$ not the time-ordering 
or $\star$-operation)
and provides the higher derivative terms 
if one evaluates the commutator of the step function 
and the $\star$-product, 
which leaves the time derivatives 
of the fields as well as of the spectral functions.
One sees the similar behavior for higher order terms,
which will be published elsewhere.
 
Second, given the S-matrix of Eq~(\ref{s-matrix}), 
the scattering amplitudes can be constructed as a 
perturbative series in the coupling constant.   
This S-matrix is obtained using the Langrangian formalism 
in the Heisenberg picture.  
The equivalence of the Hamiltonian formalism such as in \cite{gkl} 
is not easy to see 
since the symplectic structure is not 
simply tractable due to the explicit time dependence of fields 
in the interaction Langrangian.

Third, suppose one tries 
to obtain an interaction field at time $t$ 
from the in-field.
In the ordinary interaction picture 
one defines the unitary transformation,
\beq
\phi_I (t) = U(t)^\dagger \,\phi_{\rm in} (t)\, U(t) 
\label{unitary}
\eeq
with $S= \lim_{t \to \infty} U(t)$.
Requiring the dynamical evolution both  
for the in and interaction fields, 
$\dot \phi_{\rm in} (t) 
\equiv [-iL_0 (\phi_{\rm in } ), \phi_{\rm in} (t)]$ and 
$\dot \phi_I (t) \equiv [-iL (\phi_I ), \phi_I (t) ]$, 
one would obtain the dynamical equation for the unitary operator,
$\dot U (t) = i L_I (\phi_{\rm in}  (t) ) \, U (t) \,$,
on the condition that
\beq
U \, L(\phi_I)  U^\dagger = L(\phi_{\rm in})\,.
\eeq
However, this condition is not compatible with 
the Eq.~(\ref{unitary})
due to the space-time noncommutative $\star-$product 
of the action.
The unitary operator $U(t)$ does not 
transform the in-field action to interaction field action.
The same conclusion also goes for Heisenberg picture.
Nevertheless, the difficulty of constructing the unitary operator  
does not mean 
that one cannot construct S-matrix.
The transformation 
between in-field and out-field Eq.~(\ref{S-relation}) 
is enough for the existence of S-matrix 
Eq.~(\ref{s-matrix}).

\section{Propagator and Unitarity} 
\noindent
To illustrate the point described in section 2 
more concretely, we will consider  $\phi^3$ theory, 
\beq  L_I (t) 
= -\frac{g}{3!} \,\int d^{D-1}x \,\, \frac12\,
( \phi_\star^3 (x,t)+ {\rm h.c. })
\eeq
and calculate the one-loop 
contribution to the propagator in momentum space.
The mometum space calculation will be complementary  
with the coordinate space representation given
in section 2. 

The connected one loop contribution to the self-energy with 
external momentum $p_1$ and $p_2$ is given 
from the second term of S-matrix in Eq.~(\ref{s-matrix}),
denoted as $S_2$ in the following:
\bea
\langle p_1|  S_2  |p_2 \rangle_c 
= \begin{picture}(70,23)(-3,10)
\put(00,13){\line(3,0){20}}
\put(20,13){\circle{2}}
\put(40,13){\circle{2}}
\put(40,13){\line(3,0){20}}
\put(30,13){\circle{20}}
\put(-5,5){\footnotesize{$p_1$}}
\put(55,5){\footnotesize{$p_2$}}
\end{picture}
= -\frac12  \int\!\int d^D x \,
d^D y\, \langle p_1|  T_{\star}
\Big( V( \phi_{\rm in}(t_1) ) 
V(\phi_{\rm in}(t_2)) \Big) |p_2 \rangle_c  
\eea
where $ \langle \,\cdots\,\rangle_c  $ refers 
to the one-particle irreducible function.
Using the one particle representation, 
$\langle p | \phi_{\rm in}(x) |0 \rangle =N e^{ipx}$
with $N$ a proper normalization constant, 
and the integration representation of the
step function
\bea
\theta(t) = - \int_{-\infty}^\infty
\frac{d\omega}{2\pi i }\, 
\frac{e^{-i \omega t}} {\omega + i \epsilon}
\eea
we have 
\beaq
\langle p_1|  S_2  |p_2 \rangle_c 
&=&  - \left( \frac{g}{3!}\right)^2
\int\!\!\int d^D x \,d^D y 
\langle p_1|  {\cal F}_{xy}
\, \Big( \theta(x^0- y^0) 
\phi_0^3 (x) \,\phi_0^3 (y) \Big) 
|p_2 \rangle_c  \,
\label{one-loop}\\
&=& \left( \frac{g}{3!}\right)^2 
\int\!\cdots\!\int  
\frac{d^D\! x \,\, d^D\! y \,\, d^D\! k \,\,d^D\! l\,\,d\omega }
{(2\pi i)\,(2\pi)^{2D}\, (\omega + i \epsilon)}\, 
e^{i x (p_1 - k -l -\omega ) 
- i y(p_2 -k  -l-\omega )}\,
\nonumber \\ 
&&\quad \times\, 
|N|^2 \, \,\tilde \Delta_+(k) \, \tilde \Delta_+(l) 
\sum_{ \{a\}\, \{b\} } 
\cos \left(\frac{a_2 \wedge a_3 }2 \right)
\cos \left(\frac{b_2 \wedge b_3 }2 \right)
\, + \, p_1 \leftrightarrow p_2 \,.
\nonumber
\eeaq
The summation is over the set of momenta, $\{a\}$ and $\{b\}$, 
\bea
\{(a_1,a_2,a_3) \}&=& \{(p_1, -k,-l-\omega ), 
(-k, p_1, -l-\omega ), (-k,-l,p_1)\}
\\
\{(b_1,b_2,b_3)\} &=& \{ (-p_2,k ,l +\omega ), 
(k,-p_2,l+\omega ), (k,l+\omega ,-p_2), k\leftrightarrow l\}
\eea
and $ \tilde \bigtriangleup_+ (k) =
2\pi \delta (k^2 - m^2 ) \theta ( k^0)$
is the Fourier transform of the free spectral function, 
\beq 
\Delta_+ (x) = \langle 0\mid \phi_{\rm in} (x) 
\phi_{\rm in} (0) \mid 0 \rangle 
= \int \frac{d^D k}{(2\pi)^D} \,\, e^{-ikx} 
\,\,\tilde \bigtriangleup_+ (k)\,.
\eeq

Integrating over coordinates $x$ and $y$,
we are left with the
momentum representation,
\beaq
\langle p_1|  S_2  |p_2 \rangle_c 
&=& 
\begin{picture}(90,23)(-10,10)
\put(00,13){\line(3,0){20}}
\put(20,13){\circle{2}}
\put(40,13){\circle{2}}
\put(40,13){\line(3,0){20}}
\put(30,13){\circle{20}}
\put(-10,10){\footnotesize{$p_1$}}
\put(10,28){\footnotesize{$p_1- k-\omega$}}
\put(28,-8){\footnotesize{$k$}}
\put(65,10){\footnotesize{$p_2$}}
\end{picture}
=\frac{g^2}{2}  \,(2\pi)^D \,\delta^D (p_1 -p_2) \,
\int\!\!\int \frac{d^D\!k\,\, d^D\! l\,\, d\omega }
{(2\pi)^{2D}\, (2\pi i)\, (\omega + i \epsilon)}
\nonumber\\
\nonumber\\
&& \times 
 \,(2\pi)^D \,\delta^D (p_1 -k-l-\omega) \,
\,|N|^2 \,\,
\tilde \Delta_+(k)\, \tilde \Delta_+(l) \,
\cos^2 \left( \frac {p_1 \wedge l}2  \right) \,.
\label{one-loop-spectral}
\eeaq
This result shows that the external energy-momentum is manifestly 
conserved. However, the internal momentum 
need not be conserved; there appears the
spurious momentum $\omega$ in the 
internal vertex, which traces back to the 
noncommutativeness of space and time coordinates. 
One may avoid this unpleasant feature by introducing the 
retarded positive spectral function,
\bea
\theta(x^0) \Delta_+ (x)
= \int \frac{d^D k}{(2\pi)^D} \,\, e^{-ikx} 
\,\,\tilde \bigtriangleup_R (k)\,,\qquad\quad
\tilde \bigtriangleup_R (k)
= \frac{i}{2\omega_k} \frac1{(k_0- \omega_k + i\epsilon)}
\eea
where $\omega_k = \sqrt{\vec k^2 + m^2}$. \ In terms of this retarded 
function, we have Eq.~(\ref{one-loop-spectral}) as
\beaq
\langle p_1|  S_2  |p_2 \rangle_c 
=\frac{g^2}{2}  \,(2\pi)^D \,\delta^D (p_1 -p_2) \,
\int \frac{d^D\!k}{(2\pi)^{2D}}\,|N|^2 \,\,
\tilde \Delta_R(k)\, \tilde \Delta_+(p-k) \,
\cos^2 \left( \frac {p_1 \wedge k}2  \right) \,.
\eeaq

The real part of the S-matrix is given as
\beq
\label{F+}
\langle p_1|\, S_2 + S_2^\dagger \, |p_2 \rangle_c  
= - (2\pi)^D \delta^D (p_1 - p_2) \,\,F_+(p_1)
\eeq
where 
\bea
F_+(p) =  g^2 \int \frac{d^D k}{(2\pi)^D} \, |N|^2 \,
\tilde \bigtriangleup_+ (k) \,\tilde 
\bigtriangleup_+ (p_1-k)\, \,
\cos^2 (\frac{p_1\wedge k}2 )
\eea
due to the identity
$\frac1{\omega + i \epsilon}
=P \Big(\frac1{\omega }\Big)
-i \pi \delta(\omega)$.
On the other hand, $ S  S^\dagger $ of the order $g^2$ comes from the 
first term in the S-matrix Eq.~(\ref{s-matrix}) :
\beaq
\langle p_1|\, S_1  S_1^\dagger \,|p_2 \rangle_c 
&=&  \frac{g^2}{2}\, 
\int\!\cdots\!\int 
\frac{d^D x \,d^D y \, d^D k \,d^D l}{(2\pi)^{2D}} \,|N|^2 \, 
\tilde \Delta_+(k)\, \tilde \Delta_+(l) 
\nonumber\\
&& 
\quad\times\,
 e^{i x(p_1 - k -l) - i y (p_2 -k -l)  }\,
\cos^2 \left(\frac{ p_1 \wedge k}2 \right)\,
+ \, p_1 \longleftrightarrow p_2
\nonumber\\
&=& 
\,(2\pi)^D \delta^D (p_1 - p_2) \,F_+(p_1) \,.
\eeaq
This demonstrates the unitarity relation
up to the one-loop order: 
\beq
\langle p_1|\,S_2 + S_2^\dagger\,|p_2 \rangle_c 
+ \langle p_1|\, S_1  S_1^\dagger\,|p_2 \rangle_c =0\,.
\eeq
In other words, the one-loop correction $F_+ (p)$ 
is written in terms of on-shell particles only,
\beq
F_+ (p) = \sum_{\footnotesize \begin{array} {c}
l^0>0\,, l^2 =m^2 \\ k^0>0\,, k^2 =m^2 \end{array}} 
\left| 
\begin{picture}(50,23)(-3,10) 
\put(00,13){\line(3,0){20}}
\put(20,13){\circle{2}}
\put(20,13){\line(2,1) {15}}
\put(20,13){\line(2,-1){15}}
\put(0,16){\footnotesize{$p$}}
\put(40,23){\footnotesize{$k$}}
\put(40,-3){\footnotesize{$l$}}
\put(28,17){\tiny{$>$}}
\put(28, 6){\tiny{$>$}}
\end{picture}
\right|^2 \,.
\eeq
$F_+(p)$ gives a finite contribution when $p^2 > 4m^2$. 
In CM ($p^0=E, \vec p =0 $ ), this gives
\beq
F_+(p) = (4\pi)^{2-D} \frac{(E^2 - 4 m^2 )^{(D-3)/2}} { 2E}
\int d \Omega  \cos ^2(\frac {p\wedge l }2 ) \,.
\eeq

One might think that using the property of the Feynman propagator 
$\, i\! \bigtriangleup_F (x) = 
\theta (x^0) \bigtriangleup_+ (x) +
\theta (-x^0) \bigtriangleup_- (x) $;
\beq
-\Big(\!\bigtriangleup_F (x) \Big)^2 
= \theta (x^0) \Big(\!\bigtriangleup_+ (x)\Big)^2
+ \theta (-x^0) \Big(\!\bigtriangleup_- (x) \Big)^2\,,
\label{feynman-product}
\eeq 
the one-loop contribution Eq.~(\ref{one-loop}) 
can be rewritten in terms of the Feynman propagator 
instead of the spectral function used in Eq.~(\ref{one-loop-spectral}), 
\beaq
G(p)&=& 
\begin{picture}(70,23)(-10,10)
\put(00,13){\line(3,0){20}}
\put(20,13){\circle{2}}
\put(40,13){\circle{2}}
\put(40,13){\line(3,0){20}}
\put(30,13){\circle{20}}
\put(1,16){\footnotesize{$p$}}
\put(20,28){\footnotesize{$p-l$}}
\put(28,-8){\footnotesize{$l$}}
\put(55,16){\footnotesize{$p$}}
\end{picture}
\nonumber\\
&&\nonumber\\
&=& -\,\frac{g^2}4 \, \delta^D (p_1 -p_2) \,
\int\!\int  d^D k \, d^D l \, 
\delta^D (p_1 - k -l) 
\nonumber\\
&&\qquad\qquad \qquad\qquad \qquad\qquad \times \,|N|^2 \,
\tilde \Delta_F(k) \, \tilde \Delta_F(l) \,
\cos^2 \left( \frac {k \wedge l}2  \right) 
\nonumber\\
&=& 
\frac{g^2}4 \, \delta^D (p_1 -p_2) \,
\int d^D l  \,
\frac{ |N|^2 \, \cos^2 \Big(\frac{p_1 \,\wedge l }2 \Big)}
{((p-l)^2-m^2+i\varepsilon)(l^2-m^2+i\varepsilon)} \,,
\label{one-loop-feynman}
\eeaq
as has been carried out in \cite{stnc}. 
The two approaches are equivalent 
if the non-commutativeness  involves 
in the space coordinates only ($\theta^{0i} =0$).
In this case  the $\star$-operation 
and the time-ordering commutes with each other and 
Eq.~(\ref{feynman-product}) is allowed.

However, for the problematic space-time non-commutative case 
($\theta^{0i} \ne 0$),  two approaches are not the same anymore. 
In this case,
the time ordering need to be done before $\star$-operation 
and Eq.~(\ref{feynman-product}) is not justified
since
\bea
- \bigtriangleup_F (x_1 -y_1 ) \,
\bigtriangleup_F (x_2 -y_2 )\,
&\ne& \theta (x_1^0 -y_1^0) \,
\bigtriangleup_+ (x_1-y_1)\, \bigtriangleup_+ (x_2-y_2)\\
&& \qquad
+ \,\theta (-x_1^0 + y_1^0) \,
\bigtriangleup_- (x_1-y_1) \, \bigtriangleup_- (x_2-y_2) \,,\\
- \bigtriangleup_F (x_1 -y_1 ) \,
\bigtriangleup_F (x_2 -y_2 )\,
&\ne& \theta (x_1^0 -y_1^0) \,\theta (x_2^0 -y_2^0) \,
\bigtriangleup_+ (x_1-y_1)\, \bigtriangleup_+ (x_2-y_2)
\\ && 
+ \,\theta (-x_1^0 + y_1^0) \,\theta (-x_2^0 + y_2^0) \,
\bigtriangleup_- (x_1-y_1) \, \bigtriangleup_- (x_2-y_2) \,,
\eea
and there are cross terms.
Some of this step functions are ill-defined
once the $\star$-operation is performed
and the $x_i$'s ($y_i$'s) are identified as $x$ ($y$),
and some of the step functions 
provide additional contribution 
to the final result.
From this behavior, it is not surprising to see 
that the Feynamn rule will not be 
the naive generalization 
such as in Eq.~(\ref{one-loop-feynman}).
In contrast to this, 
the use of the spectral function 
$\bigtriangleup_\pm$ with the appropriate 
time-ordering takes care of the subtleties 
and results in the correct unitarity condition. 

The similar one-loop result can be used 
to check the unitarity of the scattering matrix 
in $\phi^p_\star$ theory. And one can 
perform higher loop calculation  
without any conceptual difficulty. 
We back up this idea further 
using an integrable field theory.  
In 1+1 dimension, 
non-relativistic nonlinear Schr\"odinger model is
known to be integrable 
and its exact S-matrix is known~\cite{thacker}. 
Here, we give the exact two-particle scattering matrix
for the non-commutative version of the model 
with $\theta^{01}=\theta \varepsilon ^{01}$. 
This model is the 1+1 dimensional version 
of the non-relativistic $\phi^4$ theory~\cite{bksy}.

\section{Non-relativistic nonlinear 
Schr\"odinger Model in 1+1 dimension} \noindent  

The free Lagrangian of this model is the conventional Schr\"odinger one
 and the interaction Lagrangian is given as
\beq
L_I (t)  = - \frac v4 \int d {\bf x}  \,\,  \psi^\dagger \star \psi^\dagger \star
\psi \star \psi  (t, {\bf x}) \
\eeq
where we use the bold-face letter 
for spatial vector to distiguish from the 2-vector. 
The in-field $\psi_{\rm in}$ 
satisfies the commutation relation,
$[ \psi_{\rm in} ({\bf x},t) ,
\psi^\dagger_{\rm in} ({\bf y},t)] 
= \delta ({\bf x}-{\bf y})$
and is given in momentum space,
\beq  
\psi_{\rm in}  (x)= \int \frac{d^2 k}{(2 \pi)^2} \,
\tilde D_+(k) \,
a( {\bf k})\, e^{-i k x } \,,
\,\qquad
\psi_{\rm in}^\dagger ( x)  = \int \frac{d^2 k} {(2 \pi)^2} \,
\tilde D_+(k) \,
a^\dagger  ( {\bf k})\, e^{i kx }\,,
\eeq
with $[a({\bf k}), a^\dagger ({\bf l})] 
= 2 \pi \delta ({\bf k } -{\bf l})$
and 
$\tilde D_+(p) = 2 \pi \delta (p^0 - {\bf p}^2/2 )$. 
In this non-commutative case also, 
the particle number operator 
$ {\cal N}= \int d{\bf x} \psi^\dagger \psi$ is conserved 
and this simplifies the perturbative calculation greatly.
The propagator is given 
in terms of the positive spectral function,
\beq
D_+(x) = <0| \psi_{\rm in} (x) \psi_{\rm in} ^\dagger (0)|0> = 
 \int \frac{d^2 p}{(2\pi)^2}\, e^{-ipx} \tilde D_+ (p)\,.
\eeq
The time-ordering in the S-matrix is simplified due to the 
absence of anti-particles in this 
non-relativistic case,
\beaq
D_R(x)&=&\theta(x^0)  
<0| \psi_{\rm in} (x) \psi_{\rm in} ^\dagger (0)|0>
\nonumber\\
&=&  - \int_{-\infty}^\infty
\frac{d\omega}{2\pi i }\, 
\frac{e^{-i \omega x^0}} {\omega + i \epsilon}
 \int \frac{d^2 p}{(2\pi)^2}\, e^{-ipx} \tilde D_+ (p)
= \int \frac{d^2 p}{(2\pi)^2}\, e^{-ipx} \tilde D_R(p)
\eeaq
with $\tilde D_R(p) = i/ (p^0 - {\bf p}^2/2 +i \epsilon )$.

The four point vertex is given as 
\beaq
&&\Gamma_0 (p_1, p_2; p_3,p_4)= 
\begin{picture}(60,25)(-3,10) 
\put(10,00){\line(2,1) {48}}
\put(10,23){\line(2,-1){48}}
\put(32,12){\circle{2}}
\put(00,23){\footnotesize{$p_1$}}
\put(00,00){\footnotesize{$p_2$}}
\put(60,23){\footnotesize{$p_3$}}
\put(60,00){\footnotesize{$p_4$}}
\put(15,17.5){{\tiny$>$}}
\put(15,2.5){\tiny{$>$}}
\put(45,17.5){{\tiny$>$}}
\put(45,2.2){\tiny{$>$}}
\end{picture}
\nonumber \\
\nonumber\\
&&\qquad =
- iv (2\pi)^2 \delta^2 (p_1+p_2 -p_3-p_4) 
 \cos (\frac {p_1 \wedge p_2}2 )\,
\cos (\frac {p_3 \wedge p_4}2 )\,.
\eeaq
One-loop correction to the vertex is given as
\beaq 
&&\Gamma_1 (p_1, p_2;p_3,p_4)  = 
\begin{picture}(85,30)(-3, 10)
\put(10,3){\line(2,1){20}}
\put(10,23){\line(2,-1){20}}
\put(30,13){\circle{2}}
\put(50,13){\circle{2}}
\put(50,13){\line(2,1){20}}
\put(50,13){\line(2,-1){20}}
\put(40,13){\circle{20}}
\put(0,23){\footnotesize{$p_1$}}
\put(0,00){\footnotesize{$p_2$}}
\put(15,17.2){{\tiny$>$}}
\put(15,5.5){\tiny{$>$}}
\put(60,18){{\tiny$>$}}
\put(60,4){\tiny{$>$}}
\put(30,28){\footnotesize{$p-l$}}
\put(38,-8){\footnotesize{$l$}}
\put(36,21.25){\tiny{$>$}}
\put(36,1.75){\tiny{$>$}}
\put(73,23){\footnotesize{$p_3$}}
\put(73,00){\footnotesize{$p_4$}}
\end{picture}
\nonumber\\
\nonumber\\
&&\qquad 
= -\frac {v^2}2  \, (2\pi)^2 \delta^2 (p_1+p_2 -p_3-p_4) 
\xi (p_1,p_2) \, 
\cos( \frac{p_1 \wedge p_2}2  )\,
\cos( \frac{p_3 \wedge p_4}2  ) \,,
\eeaq
where $\xi$ is defined as
\bea
\xi (p_1,p_2 )  = 
\int \frac{d^2 l}{(2\pi)^2 } \,  
\tilde D_R(l)\, \tilde D_+(p-l) \,
\cos^2 \left( \frac{l \wedge p}2 \right)
\eea
with $p= p_1 +p_2 = p_3 +p_4 $.
When $p_1$ and $p_2$ are on-shell, 
its value is given by 
\beq
\xi (p_1,p_2 ) = 
\frac 1{|{\bf p_1 } - {\bf p_2} |} \, 
\cos \left(\frac {\theta  |{\bf p_1}| 
|{\bf p_2 }| |{\bf p_1} - {\bf p_2} |}4 \right)\,
e^{\frac {i\theta  |{\bf p_1}| 
|{\bf p_2 }| |{\bf p_1} - {\bf p_2} |}4 }\,.
\eeq

Higher loop corrections are given in chained bubble diagrams 
and the complete loop corrections to 
the vertex are given in the geometric sum,
\beaq
&&\Gamma (p_1,p_2;p_3, p_4)
= \Gamma_0 (p_1,p_2;p_3, p_4) \, 
\left( 1 + \left( \frac{-i v\, \xi (p_1,p_2 )}2 \right) 
+ \left(\frac {-i v\,\xi(p_1,p_2)}2 \right)^2 \cdots \right) 
\nonumber \\
&&\qquad \quad=  (2\pi)^2 \delta^2 (p_1+p_2 -p_3-p_4) 
\cos (\frac {p_1 \wedge p_2}2 )\,
\cos (\frac {p_3 \wedge p_4}2 )\,
\frac{- iv }{ 1+  i \frac{v}2  \xi(p_1,p_2) } \,.
\eeaq
From this one obtains the on-shell 2-particle scattering amplitude,
\beaq
&&\langle p_3, p_4| S |p_1, p_2 \rangle_{(2,2)} 
= \Big(\delta({\bf p}_1 -{\bf p}_3) \,
\delta({\bf p}_2 -{\bf p}_4) 
+\delta({\bf p}_1 -{\bf p}_4) \,
\delta({\bf p}_2 -{\bf p}_3) \Big)\,\, S_{(2,2)}
\nonumber\\
&&S_{(2,2)} 
= 1 + \left( \frac{\xi(p_1,p_2) +  \xi^*(p_1,p_2)}2 \right)\,
 \left(\frac{ -i v }{ 1+  i \frac{v}2  \xi(p_1,p_2) } \right)\,
= \frac{1-i \frac{v}2 \xi^*(p_1,p_2) }{1+i \frac{v}2  \xi(p_1,p_2)} \,.
\eeaq
This exact scattering matrix is manifestly unitary,
$S_{(2,2)}^\dagger = S_{(2,2)}^{-1}\,$\/, and 
smoothly reduces to the commutative field theoretical value 
if we put the non-commutative parameter $\theta =0$.\\

To summarize, we have demonstrated  
how the perturbative analysis 
in the space-time non-commutative field theories 
respects the unitarity if S-matrix is defined 
with the proper time-ordering 
and the free spectral function is used  
instead of the Feynmann propagator.

\vskip 0.5cm
          
It is acknowledged that this work was supported in part 
by the Basic Research Program of the Korea Science 
and Engineering Foundation Grant number 
R01-1999-000-00018-0(2002) (CR)
and by Korea Research Foundation 
under project number KRF-2001-005-D00010 (JHY).

\end{document}